\documentclass[aip,twocolumn,reprint]{revtex4-1}

\usepackage{graphicx}
\usepackage{dcolumn} % Align table columns on decimal point
\usepackage{bm}      % bold math
\usepackage{amsmath,amssymb,amsthm,amsbsy}

\draft % marks overfull lines with a black rule on the right

\begin{document}


\title{Microrheological approach for the viscoelastic response of gels}

\author{L. G. Rizzi}

\affiliation{1.\,Depto.\,de\,F\'isica,\,Universidade\,Federal\,de\,Vi\c{c}osa,\,CEP:\,36570-000,\,Vi\c{c}osa-MG,\,Brazil.}

\date{\today}

\begin{abstract}
	In this paper I present a simple and self-consistent framework based on microrheology that allows one to obtain the mechanical response of viscoelastic fluids and gels from the motion of probe particles immersed on it.~By 
	considering a non-markovian Langevin equation, I obtain 
general expressions for the mean-squared displacement and the time-dependent diffusion coefficient that are directly related to the memory kernels and the response function, and which allow one to obtain estimates for the complex shear modulus and the complex viscosity of the material.
	The usefulness of such approach is demonstrated by applying it to describe experimental data on chemically cross-linked polyacrylamide through its sol-gel transition.
\end{abstract}

%\pacs{}% insert suggested PACS numbers in braces on next line

\maketitle %\maketitle must follow title, authors, abstract and \pacs

\section{Introduction}

	Gelatin is one of those curious materials that call attention of any children, as it happened with the young James Clerk Maxwell~\cite{campbellbook}, who later 
put forward one of the first descriptions of what we known as viscoelasticity~\cite{maxwell1877encbrit}.~Constitutive
	equations, like those defined by Maxwell's and Kelvin-Voigt's models~\cite{ferrybook,larsonbook}, are the basis of many theoretical approaches that are still used today to describe the relaxation modulus $G(t)$ and the compliance $J(t)$ of viscoelastic materials.~In fact,
	the mechanical response of gels can be more complicated and usually one resort to generalized models which assume that $G(t)$ can be evaluated from a relaxation time distribution $H(\lambda)$, also known as relaxation spectrum~\cite{ferrybook,mourswinter}, that is
\begin{equation}
G(t) = G_0 + \int_{0}^{\tau_c} H(\lambda) \frac{e^{-t/\lambda}}{\lambda} d\lambda ~~,
\label{relaxation_modulus}
\end{equation}
where $\lambda$ denotes the relaxation times below a cutoff value $\tau_c$ and $G_0$ corresponds to a discrete infinite-time contribution for the shear modulus.

	The basic idea is that $H(\lambda)$ should include informations about all the microscopic and mesoscopic structures of the gel~\cite{mourswinter}, and that would provide some reasoning on what is been observed in the complex shear modulus, $G^{*}(\omega)=G^{\prime}(\omega)  + i G^{\prime \prime}(\omega)$.
	The storage, $G^{\prime}(\omega)$, and the loss, $G^{\prime \prime}(\omega)$, moduli are the main rheological properties measured in oscillatory sweeping experiments~\cite{larsonbook}
and can be obtained from $G(t)$ through a Fourier transform~\cite{doibook}.
	In particular, when the gel network is not formed yet, one have that $G_0=0\,$Pa and both $G^{\prime}(\omega)$ and $G^{\prime \prime}(\omega)$ goes to zero at low frequencies, 
and that is a characteristic behavior of liquid-like solutions.~On
	 the other hand, 
the material will behave like a semisolid~\cite{raobook} if the storage modulus of a gel is given by $G^{\prime}(\omega) = G_0$ at low frequencies, and that is usually interpreted in terms of the formation of a network.
	Unfortunately, the evaluation the complex modulus $G^{*}(\omega)$ from $H(\lambda)$ is not a trivial task and in the most of cases it is done only numerically~\cite{ferrybook,mourswinter}.

	Although several descriptions of gelation are possible~\cite{zaccarelli2007jphyscondmatt}, most of the successful theoretical developments were based on the analogy between percolation and the sol-gel transition~\cite{florybook,degennes,rubinsteinbook}, including the pioneering works done by Flory~\cite{flory1941jacs} and Stockmayer~\cite{stockmayer1943jcp}.~Fractality 
	was a concept that emerged naturally from the percolation analogy, and Ref.~\cite{martin1991review} present a review on that topic, including complementary approaches that are based on the aggregation kinetics.~For example,
	in a recent study presented in Ref.~\cite{zaccone2014jrheol}, the authors established an interesting link between the growth kinetics mechanisms and the viscoelastic response of the solution by assuming that the relaxation times of the gels structures are related to the size of the clusters in solution.~Their
	approach lead to a semi-empirical expression for the relaxation spectrum, {\it i.e.},~$H(\lambda) \propto \lambda^{-q} \exp[-(\lambda/\tau_c)^{w}]$,
with the exponents $q$ and $w$ related to the fractal dimension of the clusters, and that was used with Eq.~\ref{relaxation_modulus} to estimate the shear moduli of the colloidal suspensions~\cite{zaccone2014jrheol}.

	Fractal properties of gels are usually probed by light scattering experiments and, in terms of dynamics, it is worth mentioning the work of Krall and co-workers~\cite{krall1997physA,krall1998prl}, which have related the measurements of the dynamic structure factor $f(q,t)$ to an expression for the mean-squared displacement of the fractal structures present in colloidal gels.~Importantly, 
	they found a function $f(q,t)$ that decays to a finite plateau at long times, and interpreted that in terms of a finite elastic modulus $G_0$.

	A coupled of years before, Mason and Weitz~\cite{masonweitz1995prl}, which are considered the founders of microrheology~\cite{waigh2005review,waigh2016review},
established a generalized Stokes-Einstein relationship (GSER) and, by exploring light scattering experiments as well, demonstrate that one can extract the complex modulus $G^{*}(\omega)$ of colloidal systems from the mean-squared displacement (MSD) $\langle \Delta r^{2}(t) \rangle$ of probe particles.~At
	 the linear viscoelastic (LVE) response regime~\cite{rizzi2018eac}, one can use such GSER to estimate the compliance $J(t)$ of the viscoelastic material from the MSD of probe particles immersed on it as~\cite{squires2010annrev}
\begin{equation}
J(t) = \frac{3 \pi a}{d k_BT}  \langle \Delta r^{2}(t) \rangle ~~,
\label{compliance_MSD}
\end{equation}
where $a$ is the radius of the probe particles, $k_B$ and $T$ are the Boltzmann's constant and the absolute temperature, respectively, and $d$ is the number of degrees of freedom of the random walk ({\it e.g.},~$d=3$ for diffusing wave spectroscopy, and $d=2$ for particle tracking videomicroscopy~\cite{waigh2016review,rizzi2018eac}).

	There are a few exact interrelations between the rheological functions~\cite{ferrybook}, {\it e.g.},~between the complex modulus and the complex viscosity,~$G^{*}(\omega) = i \omega \, \eta^{*}(\omega)$,
and one can conveniently use Eq.~\ref{compliance_MSD} to evaluate the complex shear modulus by considering~\cite{rizzi2018eac}
\begin{equation}
G^{*}(\omega) = \frac{1}{i \omega \hat{J}(\omega)} ~~,
\label{complex_modulus_compliance}
\end{equation}
where $\hat{J}(\omega)$ is the 
Fourier transform of the compliance $J(t)$, which might be evaluated exactly through a Fourier-Laplace integral, or numerically by the method proposed in Ref.~\cite{manlio2009pre}.
	Equations~\ref{compliance_MSD} and~\ref{complex_modulus_compliance} opened the way for many techniques that were used to measure the MSD of probe particles to be applied in the characterization of viscoelastic materials~\cite{rizzi2018eac}.
	In particular, there are several studies in the literature that use particle tracking microrheology to obtain the compliance and shear moduli of protein systems during the gelation transition~\cite{furst2008prl,larsen2008korea,corrigan2009langmuir,corrigan2009eurphysJE,donald2014langmuir}.

	In this paper I consider an approach based on Langevin dynamics to derive general expressions for the MSD $\langle \Delta r^{2}(t) \rangle$ and the time-dependent diffusion coefficient $D(t)$ of particles immersed in a viscoelastic material described by effective elastic and friction constants.~I 
	provide explicity expressions that link the MSD to the response and memory functions and discuss how such Langevin equation is related to a Fokker-Planck type of equation.~By
	using a microrheological approach based on Eqs.~\ref{compliance_MSD} and~\ref{complex_modulus_compliance}, I obtain the complex shear moduli $G^{*}(\omega)$ and the complex viscosity $\eta^{*}(\omega)$ of the material, and show how such rheological properties are connected to the effective constants defined in the Langevin equation.
	I validate my approach by considering experimental data on the sol-gel transition of chemically cross-linked polyacrylamide gels~\cite{furst2008prl}, and demonstrate that the expression for the MSD used to describe the dynamics of probe particles in a gel phase of fibrillar proteins~\cite{rizzi2017jcp} can be also used to describe their dynamics in the sol phase.

\section{Microrheological approach}
\label{micro_approach}

\subsection{Langevin equation}

	First, consider that the motion of probe particles immersed in a viscoelastic material can be described by an overdamped Langevin equation that is given by
\begin{equation}
\zeta \frac{d \vec{r}}{dt} = - \kappa \int_{0}^{t} dt' b(t - t') \vec{r}(t') + \vec{f}(t)~~,
\label{GLE}
\end{equation}
where $\zeta$ and $\kappa$ are the effective friction coefficient and the effective elastic constant of the material, respectively; $b(t-t')$ is a memory kernel function which indicates that the particle is submitted to a non-markovian harmonic-like potential;
$\vec{f}(t)$ is a random force with gaussian statistical properties such that $\langle \vec{f}\, \rangle = \vec{0}\,$, and 
\begin{equation}
\langle \vec{f}(t) \cdot \vec{f}(t') \rangle =  d k_B T \, \zeta  \, b(t-t') ~~.
\label{FDT_for_ft}
\end{equation}

	In order to obtain a general expression for the mean-squared displacement of the probe particles, one can apply the Laplace transform, here denoted by $\tilde{z}(t)=\mathcal{L}[\vec{z}(t);s]=\int_{0}^{\infty} \vec{z}(t) \, e^{-st}dt$, to both sides of Eq.~\ref{GLE}, which leads to
\begin{equation}
\tilde{r}(s) = \bar{\chi}(s) [ \vec{r}(0) + \zeta^{-1} \tilde{f}(s) ] ~~,
\label{laplace_transformed}
\end{equation}
where 
\begin{equation}
\bar{\chi}(s)=  \frac{1}{ s + (\kappa/\zeta) \bar{b}(s) } ~~, 
\label{response_function_LT}
\end{equation}
with $\bar{b}(s) = \mathcal{L}[b(t);s]$.~Hence, 
	the general solution of Eq.~\ref{GLE} is given by
\begin{equation}
\vec{r}(t) = \chi(t) \vec{r}(0) + \zeta^{-1} \int_{0}^{t} \chi(t-t') \vec
{f}(t') dt'  ~~,
\label{rt_solution}
\end{equation}
where $\chi(t)$ is a response function related to the memory kernel $b(t)$ through
$\chi(t) = \mathcal{L}^{-1}\{[s + (\kappa/\zeta) \bar{b}(s)]^{-1};t\}$ (see Eq.~\ref{response_function_LT}).
	Now, following Ref.~\cite{adelman1976jcp}, one can define
\begin{equation}
\vec{y}(t) = \vec{r}(t) - \chi(t) \vec{r}(0) = \zeta^{-1} \int_{0}^{t} \chi(t-t') \vec
{f}(t') dt' 
\label{vecyt}
\end{equation}
and, by considering Eq.~\ref{FDT_for_ft}, evaluate the fluctuation function of $\vec{y}(t)$, {\it i.e.},~$A(t) = \langle \vec{y}(t) \cdot \vec{y}(t) \rangle$, which is given by
\begin{equation}
A(t) = \frac{d k_B T}{\zeta} \int_{0}^{t}  \int_{0}^{t} \chi(t-t') 
b(t'-t'')
 \chi(t-t'')  dt' dt'' ~~.
\label{Afluct}
\end{equation}
	Also, one should compute its time derivative,
\begin{equation}
\frac{\partial A(t) }{\partial t}
= \frac{ 2 d k_B T}{\zeta} \chi(t) \int_{0}^{t}  
b(t-t') \chi(t') dt' ~~.
\label{Adotchib}
\end{equation}
	Now, from Eq.~\ref{response_function_LT} one have that
\begin{equation}
s \bar{\chi}(s) - 1  = - \frac{\kappa}{\zeta} \bar{b}(s) \bar{\chi}(s) ~~,
\label{deriv_chis}
\end{equation}
and, since Eq.~\ref{rt_solution} requires that $\chi(0)=1$, the inverse Laplace transform of Eq.~\ref{deriv_chis} will be given by
\begin{equation}
- \frac{\zeta}{\kappa} 
\frac{\partial \chi(t) }{\partial t}
= \mathcal{L}^{-1}[\bar{b}(s) \bar{\chi}(s);t]
= \int_{0}^{t}  
b(t-t') \chi(t') dt' ~~,
\label{invLTderivchis}
\end{equation}
so that Eq.~\ref{Adotchib} can be rewritten as
\begin{equation}
\frac{\partial A(t) }{\partial t}
= - \frac{ 2 d k_B T}{\kappa} \chi(t) 
\frac{\partial \chi(t) }{\partial t}
~~.
\label{Adot}
\end{equation}
Thus, from the above equation one can infer that the fluctuation function defined by Eq.~\ref{Afluct} should be given by
\begin{equation}
A(t) = \frac{ d k_B T}{\kappa} \left[ 1 -  \chi^{2}(t) \right] ~~,
\label{Achi}
\end{equation}
from where one can obtain the mean-squared displacement, $\langle \Delta r^2(t) \rangle \equiv A(t)$, by choosing, without loss of generality, $\vec{r}(0)=\vec{0}\,$.

\subsection{Fokker-Planck equation}
\label{fokkerplanckEQ}

	Importantly, if the random process defined by Eq.~\ref{GLE} is gaussian, one can assume that the time-dependent position distribution function is given by
\begin{equation}
P(\vec{r},t) \propto
\frac{1}{\left[ 2\pi A(t) \right]^{ d/2  }  } 
\exp
\left\{
- \frac{[\vec{r}(t)]^2}{2 A(t)}
\right\} ~~,
\label{pdistfunc}
\end{equation}
which corresponds to the solution of a Fokker-Planck type of equation that is characteristic of non-markovian processes and can be written as~\cite{adelman1976jcp}
\begin{equation}
\frac{\partial P(\vec{r},t)}{\partial t}
=
k_B T \beta(t)  \nabla_{r} \left[ \nabla_{r} P(\vec{r},t) + P(\vec{r},t) \left( \frac{ \kappa \vec{r} }{k_B T} \right) \right] ~,
\label{fokker_planck_eq}
\end{equation}
where $\beta(t)$ is a memory function defined as
\begin{equation}
\beta(t) = - \frac{1}{\kappa}  \frac{1}{\chi(t)}
\frac{\partial \chi(t) }{\partial t}
 ~~.
\label{beta_chi_relationship}
\end{equation}

	Interestingly, from Eqs.~\ref{invLTderivchis} and~\ref{beta_chi_relationship} one finds that
\begin{equation}
\beta(t) \, \chi(t)  = \frac{1}{\zeta} \int_{0}^{t} b(t-t') \chi(t') dt' ~~,
\label{beta_chi_b_relationship}
\end{equation}
which indicates that, in general, there is no simple relationship between
the memory kernel $b(t)$ defined in Eq.~\ref{GLE} and the functions 
$\chi(t)$ and $\beta(t)$.
	In particular, one can observe that the memory function $\beta(t)$ should be related, but it is not generally equal, to the mobility function $\mu(t)= ( 2d k_BT )^{-1} (\partial A(t) /\partial t )$, which
can be evaluated from Eqs.~\ref{Adot} and~\ref{beta_chi_relationship} as
\begin{equation}
\mu(t) = \beta(t) \, \chi^2(t) ~~.
\label{mobility}
\end{equation}

\subsection{Alternative Langevin equation}
\label{alternative_GLE}

	It is worth mentioning that the aforementioned Langevin approach defined by Eq.~\ref{GLE} can be written in terms of an alternative Langevin equation,
where the effective elastic force is equivalent to a delayed viscous force, that is
\begin{equation}
\kappa \vec{r}(t) = - \zeta \int_{0}^{t} u(t-t') \vec{v}(t') dt' + \vec{g}(t)
\label{GLE_alt}
\end{equation}
with $\langle \vec{g}(t) \, \rangle = \vec{0}$ and $\langle \vec{g}(t) \cdot \vec{g}(t')  \rangle = d k_B T \, \kappa \, u(t-t')$.
	One can show that the Laplace transform of Eq.~\ref{GLE_alt} yields 
\begin{equation}
\tilde{r}(s)  = \left[ s  +  \frac{(\kappa/\zeta)}{\bar{u}(s)}  \right]^{-1}
\left[   \vec{r}(0)  + \zeta^{-1} \frac{ \tilde{g}(s) }{\bar{u}(s)} \right] ~~.
\label{GLE_alt_LT}
\end{equation}
	Thus, by directly comparing Eqs.~\ref{laplace_transformed} and~\ref{GLE_alt_LT}, and identifying the first multiplying term in the above equation as $\bar{\chi}(s)$ defined by Eq.~\ref{response_function_LT}, one can verify that the two Langevin equations will be equivalent when 
$\tilde{g}(s) = \bar{u}(s) \tilde{f}(s)$, 
 and
\begin{equation}
\bar{u}(s) \, \bar{b}(s) = 1 ~~.
\label{abs_equivalence}
\end{equation}
	Interestingly, a similar result was found in 
Refs.~\cite{panja2010jstatmech_l02001,panja2010jstatmech_p06011}, where the author
obtain approximate expressions for the memory kernels of a tagged particle attached by springs to the middle of a Rouse polymeric chain.

	Once that the two approaches are equivalent, the same reasoning presented in Sec.~\ref{fokkerplanckEQ} should be valid.
	In particular, by considering Eqs.~\ref{deriv_chis} and~\ref{abs_equivalence}, one finds that
\begin{equation}
\chi(t) = - \frac{\zeta}{\kappa} \int_{0}^{t}u(t-t') \dot{\chi}(t') dt' ~~,
\label{chit_ut_dotchi}
\end{equation}
which is complementary to Eq.~\ref{beta_chi_b_relationship}.

\subsection{Mean-squared displacement (MSD)}

	Importantly, Eq.~\ref{beta_chi_relationship} sets a differential equation for
$\chi(t)$, which one can assume that have a solution that is given by
\begin{equation}
\chi(t) = \chi(0) \, e^{-\kappa \psi(t)} ~~,
\label{chit_exppsi}
\end{equation}
from where one can identify $\beta(t)= \partial \psi(t) / \partial t$, or
$\psi(t) = \int_{0}^{t} \beta(t') dt'$, with
\begin{equation}
\psi(t) = \frac{1}{\kappa} \ln \left[ \frac{\chi(0)}{\chi(t) }\right]~~.
\label{psit_chit}
\end{equation}
	Thus, the MSD can be computed from Eqs.~\ref{Achi} and~\ref{chit_exppsi} as
\begin{equation}
\langle \Delta r^2(t) \rangle = \frac{d k_B T}{\kappa} \left[ 1 - e^{-2\kappa \psi(t)} \right] ~~.
\label{MSDpsit}
\end{equation}
	In principle, such expression for the MSD can be also derived from the Fokker-Planck equation, Eq.~\ref{fokker_planck_eq}, {\it e.g.},~by considering the time derivative of $\langle \vec{r}^{\,2}(t) \rangle = \int P(\vec{r},t) r^{2} dr$ with $P(\vec{r},t)$ given by Eq.~\ref{pdistfunc} (see, {\it e.g.},~Ref.~\cite{holek2014jphyschemB}).

	In addition, one can evaluate the time-dependent diffusion coefficient as the time derivative of the MSD, that is, $D(t) = (2d)^{-1}  \partial \langle \Delta r^2(t) \rangle /  \partial t$,
which yields
\begin{equation}
D(t) = k_B T \mu(t) ~~,
\label{Dtmut}
\end{equation}
with the mobility $\mu(t)$ computed by Eq.~\ref{mobility}.

\subsection{Examples}
\label{examples}

	In this Section, I include four paradigmatic examples that are discussed in terms of the aforementioned framework and which will be useful to the following theoretical developments.

\subsubsection{Simple viscoelastic medium}
\label{viscoelastic_medium}

	The most trivial example corresponds to the case where the probe particle is subjected to a harmonic potential described by a markovian process so that the memory kernel in Eq.~\ref{GLE} is given by a delta function, {\it i.e.}, $b(t)=\delta(t)$, then $\beta=1/\zeta$ follows from Eq.~\ref{beta_chi_b_relationship},
and $\psi(t) = \int_{0}^{t}\beta(t')dt' = t/\zeta$, hence Eq.~\ref{chit_exppsi} gives $\chi(t)=e^{-\kappa t/\zeta}$.
	In this case, the MSD obtained by Eq.~\ref{MSDpsit} can be used to evaluate the compliance through Eq.~\ref{compliance_MSD}, which yields
$J(t)=(3\pi a/\kappa) [1 - \exp(-2t/\tau) ]$, from where one can identify
$\tau=\zeta/\kappa$ as a characteristic relaxation time. 
	As shown in Ref.~\cite{azevedo2020jconfser}, one can plug $J(t)$ into Eq.~\ref{complex_modulus_compliance} to obtain the storage modulus, $G^{\prime}(\omega)=(\kappa/3 \pi a)$, and the loss modulus, $G^{\prime \prime}(\omega)=(\zeta / 6 \pi a) \omega$, which are the typical response functions observed for a Kelvin-Voigt model~\cite{ferrybook}.
	Also, one can obtain the viscosity $\eta^{\prime}(\omega)= G^{\prime \prime}(\omega)/\omega = (\zeta / 6 \pi a)$, which is equivalent to the well-known Stokes' result, $\zeta = 6 \pi a \eta$.

\subsubsection{Normal diffusion in a simple fluid}

	Another example include the case where the probe particle is immersed in a simple fluid with viscosity $\eta$ and display a normal diffusion behavior. 
	Such regime can be retrieved by choosing a negative 
value for the elastic constant, $\kappa=-\zeta/(2\tau)$,
and $\chi(t) = [(t/\tau) + 1]^{1/2}$
(which is similar to set $\chi(t)\approx e^{t/2\tau}$ for $t \ll \tau$), then 
Eq.~\ref{psit_chit} yields $\psi(t)=(\tau/\zeta) \ln[(t/\tau) + 1]$ and Eq.~\ref{beta_chi_relationship} leads to $\beta(t)=\zeta^{-1} [(t/\tau) + 1]^{-1}$. 
	In this case, Eq.~\ref{mobility} yields a constant mobility function, $\mu(t)=1/\zeta$, so that $D(t)=k_BT/\zeta$, and Eq.~\ref{MSDpsit} leads to $\langle \Delta r^2 (t) \rangle = 2 d (k_BT/\zeta) t$.
	Then, one can use such MSD to obtain the mechanical response of the fluid through Eqs.~\ref{compliance_MSD} and~\ref{complex_modulus_compliance}, that is, $G^{\prime}(\omega)=0$, $G^{\prime \prime}(\omega)=\eta \omega$, and $\eta^{\prime}(\omega)=\eta$.
	It might be surprising that Eq.~\ref{GLE} could describe a simple fluid by an effective elastic media, however, those results can be also interpreted in terms of the alternative Langevin equation presented in Sec.~\ref{alternative_GLE}.
	By considering Eq.~\ref{chit_ut_dotchi}, one can check that, in this case, the effective elastic force defined in Eq.~\ref{GLE_alt} should correspond to a delayed viscous force with a kernel function given by $u(t - t')= \delta(t - t')[(t'/\tau) + 1]$.

\subsubsection{Polymer solutions}
\label{polymer_solutions}

	In general, the diffusion of probe particles immersed in solutions of polymers are characterized by a subdiffusive behavior with $\langle \Delta r^2(t) \rangle \propto t^n$ at intermediary times, 
and a normal diffusion behavior with $\langle \Delta r^2(t) \rangle \propto t$, at times longer
than the longest relaxation time of the chains~\cite{doiedwards}.
	The exponent $n$ depends on the kind of polymer considered, {\it e.g.},~$n=1/2$ for flexible polymeric chains described by the Rouse model, and $n=3/4$ for semiflexible polymers~\cite{rubinsteinbook}.
	Based on Refs.~\cite{panja2010jstatmech_p06011,panja2010jstatmech_l02001}, the dynamics of particles subjected to polymeric chains could be attained, in principle, by considering a kernel function like $u(t) \propto \, t^{-n} e^{-t/\tau}$.
	Interestingly, the kernel function $u(t)$ for $n=1/2$ presented in Refs.~\cite{panja2010jstatmech_p06011,panja2010jstatmech_l02001} has the same functional form of the approximated expression found for the relaxation modulus $G(t)$ in Ref.~\cite{rubinsteinbook}
when obtaining $G^{*}(\omega)$ for the Rouse chains.
	In principle, one could compute the Laplace transform of $u(t)$
and consider Eqs.~\ref{response_function_LT} and~\ref{abs_equivalence} to obtain $\bar{\chi}(s)$,
however, even for such approximated expression, the response function does not have a simple inverse Laplace transform.~Anyhow, 
it is instructive to consider times shorter than $\tau$,
where the MSD is given by $\langle \Delta r^2(t) \rangle \approx (d k_BT/\kappa) (t/\tau)^{n}$, then 
Eq.~\ref{compliance_MSD} yields $J(t) \approx (3 \pi a /\kappa) (t/\tau)^{n}$
so that $\tilde{J}(s) = \mathcal{L}[J(t);s]= 
%3 \pi a (\alpha/\kappa) \tau^{-n} \mathcal{L}[t^{n},s]= 
 (3 \pi a/\kappa\tau^{n})  \Gamma(n+1)/s^{n+1}$, where
$\Gamma(n+1)$ is the usual gamma function.~Thus, 
	by replacing $s=i\omega$ and using Eq.~\ref{complex_modulus_compliance},
one finds that 
$G^{\prime}(\omega)$ and $G^{\prime \prime}(\omega)$ will be both proportional to $\omega^{n}$ at frequencies above $\tau^{-1}$.
	For instance, $G^{\prime}(\omega) \propto G^{\prime \prime}(\omega) \propto \omega^{1/2}$ for Rouse chains, and 
$G^{\prime}(\omega) \propto G^{\prime \prime}(\omega) \propto \omega^{3/4}$ for semiflexible chains.~As 
	discussed in Ref.~\cite{leonam2020jnonnew}, solutions with structures of intermediary size and/or flexibility might display intermediary values of $n$.

\subsubsection{Polymer network}

	Finally, it is worth mentioning the widespread expression that have been used in the literature to describe the motion of probe particles in viscoelastic materials, which is given by
\begin{equation}
\langle \Delta r^2(t) \rangle = \frac{d k_B T}{\kappa} \left[ 1 - e^{-(t/\tau)^{n}} \right] ~~.
\label{MSDkrallweitz}
\end{equation}
	To my knowledge, such expression was first derived by Krall and co-workers~\cite{krall1997physA,krall1998prl} by assuming a fractal structure for the gel network using ideas that are closely related to a description based on percolation theory.
	Interestingly, one can obtain Eq.~\ref{MSDkrallweitz} from Eq.~\ref{MSDpsit} by choosing $\chi(t) = e^{-(1/2) (t/\tau)^n}$, so that Eq.~\ref{psit_chit} leads to $\psi(t) = (2\kappa)^{-1} (t/\tau)^{n}$ and a memory function given by $\beta(t)=\partial \psi(t)/\partial t=n (2 \kappa \tau)^{-1} (t/\tau)^{n-1}$.
	Unfortunately, as discussed in Ref.~\cite{rizzi2017jcp}, the derivative of Eq.~\ref{MSDkrallweitz} does not provide a consistent description between 
the evaluated $D(t)$ and the time-dependent diffusion coefficient observed in light scattering experiments~\cite{teixeira2007jphyschemB}; and the MSD given by Eq.~\ref{MSDkrallweitz}
does not fit master curves very well~\cite{romer2014epl,rizzi2017jcp}.
	Also, the memory fuction $\beta(t)$ at long times does not decay as $t^{-1}$, so that it is not consistent with fluctuating hydrodynamics theory~\cite{adelman1976jcp}.
	In any case, probably due to its simplicity, such expression have been largely used 
in the literature (see,~{\it e.g.},~Refs.~\cite{romer2000prl,holek2006jcp,teixeira2007jphyschemB,romer2014epl,holek2014jphyschemB,calzolari2017jrheol}).

\section{Results}

	In the following I apply the framework developed in Sec.~\ref{micro_approach} 
in order to obtain self-consistent expressions which can be used to describe the dynamics
of probe particles during gelation, hence the viscoelasticity of the material, for both 
sol and gel phases.
	Also, I present a comparison with experimental data
in order to validate my approach.

\subsection{Generalized response function}

	First, I consider that the viscoelastic material present a local response function which is similar to that discussed in the last example of Sec.~\ref{examples}, that is, 
$\chi_{\varepsilon}(t) = e^{-\varepsilon \psi_{\kappa}(t)}$, 
where $\psi_{\kappa}(t) = \alpha (2\kappa)^{-1} (t/\tau)^n$ and 
$\varepsilon$
 is a local effective elastic constant which might depend generally on the local properties of the structures in the viscoelastic material, {\it e.g.}, cluster/chain sizes in solutions, and mesh sizes/coordination numbers in networks.~In
	this case, the exponent $n$ in $\psi_{\kappa}(t)$ is directly related to a structural dynamic exponent, {\it e.g.},~$n=1/2$ for Rouse chains and $n=3/4$ for semiflexible polymers~\cite{leonam2020jnonnew}, and it should characterize the exponent observed at the gelation transition; and $\alpha$ is an exponent that characterizes the distribution of elastic constants 
$\varepsilon$, 
which it is assumed to be given by a gamma distribution~\cite{crooksbook}, that is
\begin{equation}
\rho_{\alpha}(\varepsilon) = 
\frac{\alpha^2}{4 |\kappa| \, \Gamma( 1 + \alpha/2)}
\left( \frac{\alpha}{2 \kappa} \varepsilon \right)^{-(1 -\alpha/2)}
\exp \left( - \frac{\alpha}{2\kappa}  \varepsilon  \right)
~~,
\label{k_distribution}
\end{equation}
where $\Gamma(1 + \alpha/2)$ denotes the usual gamma function.
	Importantly, Eq.~\ref{k_distribution} is chosen in order to have the mean value of the local elastic 
constants $\varepsilon$ consistent with $\bar{\varepsilon}=\kappa$.
	Also, it is worth mentioning that the effective elastic constants have been related to, {\it e.g.},~the size of clusters of particles/polymers with a given size~\cite{krall1997physA,dinsmore2006prl}, and that gamma distributions have been already used to describe size distributions~\cite{xueradford2009proteng}.

	Now, by changing the variable to 
$\xi=(\alpha \varepsilon/ 2 \kappa)$, so that $\chi_{\xi}(t) = e^{-\xi (t/\tau)^n }$, 
one finds that the distribution of effective dimensionless elastic constants, $\xi$, can be written as
$\rho_{\alpha}(\xi) = \xi^{-(1 -\alpha/2)}  e^{-\xi} / \Gamma(\alpha/2)$, so that the effective response function of the viscoelastic material can be evaluated exactly~\cite{gradshteynbook}, and is given by
\begin{equation}
\chi(t) = \int_{0}^{\infty} \rho_{\alpha}(\xi) \, \chi_{\xi}(t) \, d\xi = \left[ \left( \frac{t}{\tau} \right)^n + 1 \right]^{-\alpha/2} ~~,
\label{generalized_chit}
\end{equation}
which is valid for $\alpha > -2$, and the mean value of $\xi$ is related to the exponent as $\bar{\xi}=\alpha/2$.
	Note that, if $\alpha$ is negative, one should also assume that $\kappa<0$, which leads to negative values of $\bar{\xi}$ and 
$\bar{\varepsilon}$ 
as well.

	Also, by considering Eq.~\ref{psit_chit}, Eq.~\ref{generalized_chit} leads to
\begin{equation}
\psi(t) = 
 \frac{\alpha}{2\kappa}
\ln \left[  \left(  \frac{t}{\tau} \right)^{n} + 1 \right] ~~,
\label{psit}
\end{equation}
with a memory function $\beta(t)=\partial \psi(t) / \partial t$ given by
\begin{equation}
\beta(t)= \frac{\alpha n }{2 \kappa \tau} \left( \frac{t}{\tau} \right)^{n-1}
\bigg/ ~
\left[ \left( \frac{t}{\tau} \right)^{n} + 1 \right]   ~~,
\label{betat}
\end{equation}

\begin{figure}[!t]
\centering
\includegraphics[width=0.44\textwidth]{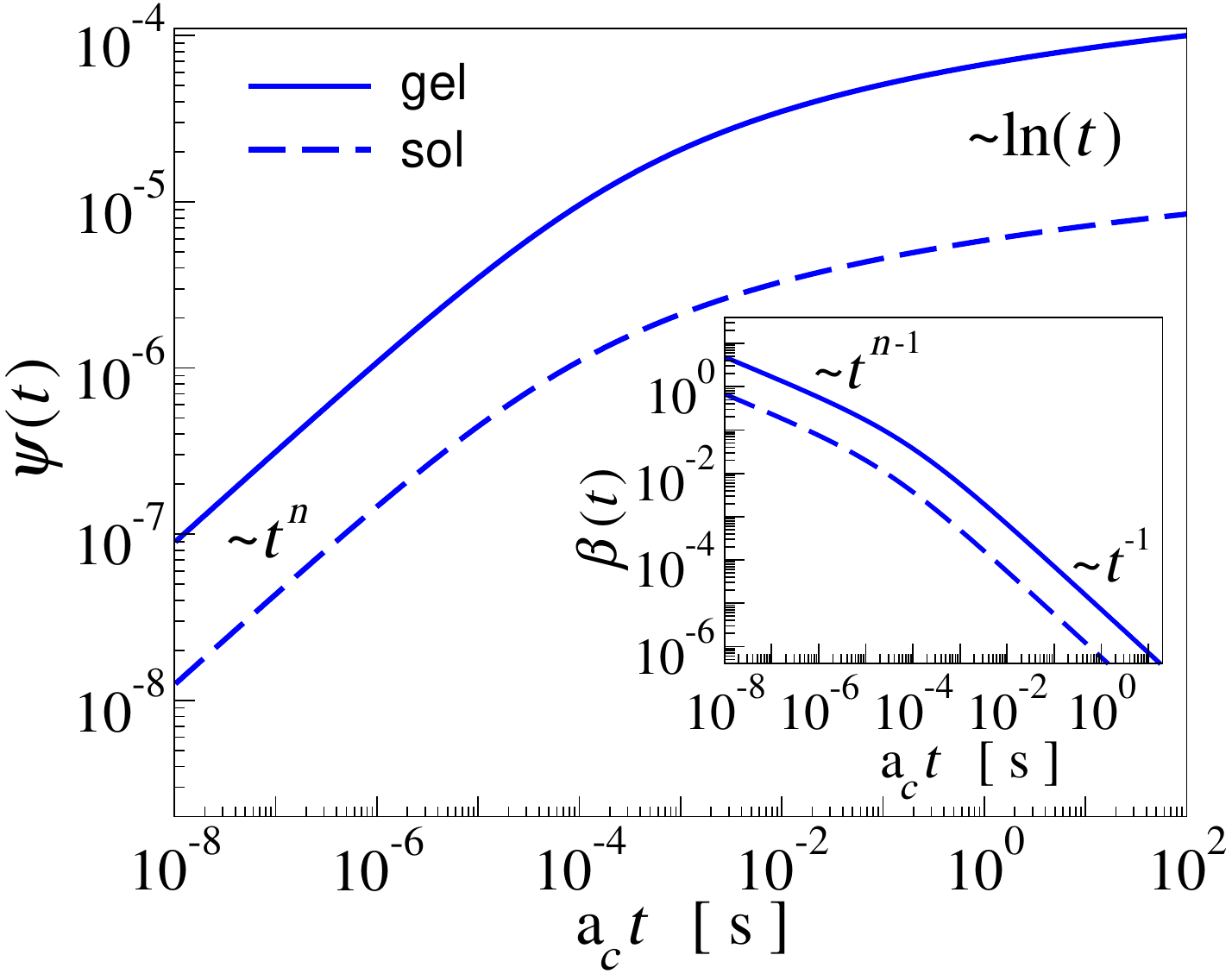}
\caption{
	Function $\psi(t)$ (main panel), Eq.~\ref{psit}, and the memory function $\beta(t)$ (inset), Eq.~\ref{betat}, that correspond to the master curves presented in Fig.~\ref{MSD}.
%	({\it i.e.},~values and parameters are given in dimensionless units).
	The dynamics of both sol and gel phases are described by the same exponent $n=0.55$
and those functions display curves with a similar behavior in a log-log plot.~For
	the gel phase one have
$\kappa=2.74 \times 10^{4}\,$Pa.nm,
$\tau=8.63 \times 10^{-5}\,$s,
and $\alpha = 0.71$, so that $\alpha n \approx 0.39$.
	For the sol phase one have that
$\kappa=-8.81 \times  10^{5}\,$Pa.nm,
$\tau=3 \times 10^{-5}\,$s,
and $\alpha=-1.8$, so that $-\alpha n \approx 1$.
}
\label{psi_beta}
\end{figure}

\begin{figure*}[!t]
\centering
\includegraphics[width=0.88\textwidth]{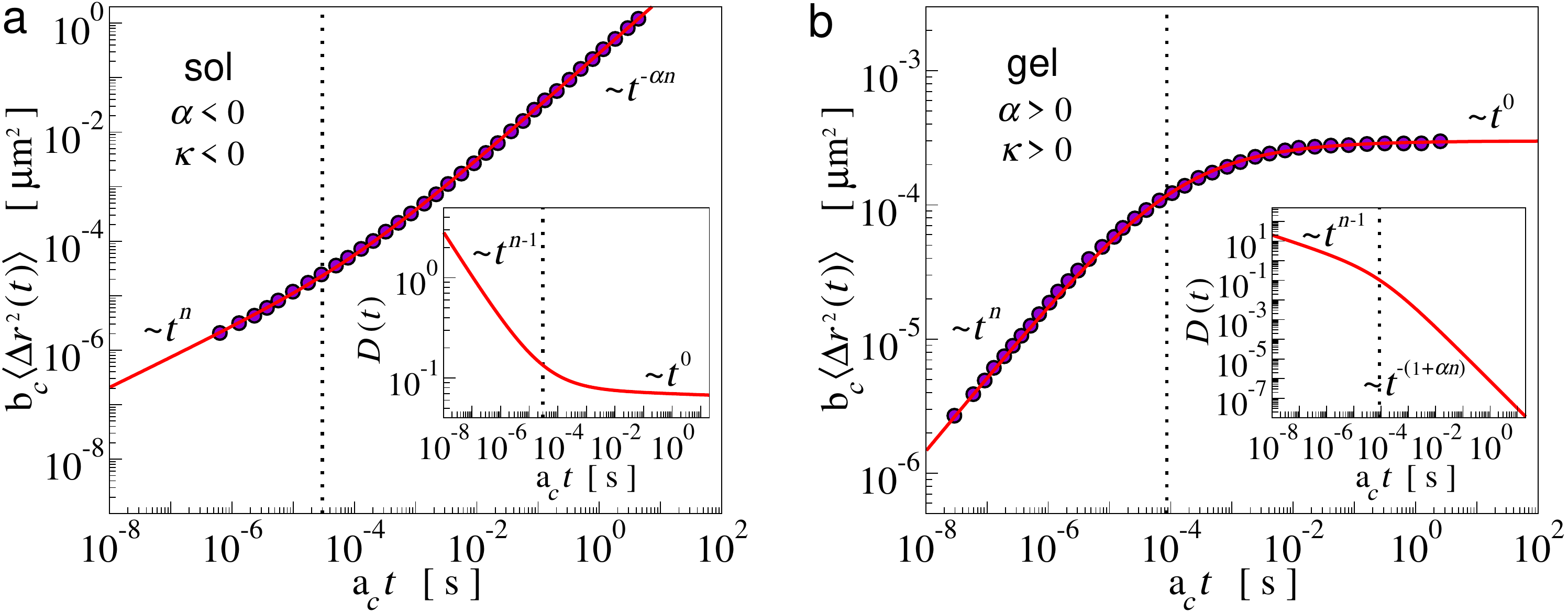}
\caption{Master curves of the mean-squared displacement illustrating the typical diffusing behaviors observed in (a)~sol and (b)~gel phases, respectively.
All parameters and curves are displayed as %dimensionless
rescaled
 quantities as they were multiplied by factors $\text{a}_c$ and/or $\text{b}_c$, {\it e.g.},~MSD $\text{b}_c\langle \Delta r^2(t) \rangle$, and time $\text{a}_c t$, which are experimentally obtained shift factors used to construct the rescaled curves.
Filled symbols correpond to the experimental data extracted from Ref.~\cite{furst2008prl} on cross-linked polyacrylamide solutions, while continuous lines denote curves obtained using~Eq.~\ref{rizzi_solution}.
Vertical dotted lines indicate the values of characteristic 
%(dimensionless) 
times, $\tau=3 \times 10^{-5}\,$s (sol phase) and $\tau=8.63 \times 10^{-5}\,$s (gel phase), which separate the subdiffusive behavior with exponent $n=0.55$ at short times from the later time regimes, in which the sol-phase is characterized by the presence of a normal diffusion behavior obtained with $\alpha=-1.8$ so that $-\alpha n \approx 1$, while the limited diffusion observed in the gel-phase is obtained with $\alpha = 0.71$.
The 
%(dimensionless)
 values of the effective elastic constant are $\kappa=-8.81 \times 10^{5}\,$Pa.nm (sol phase) and $\kappa=2.74 \times 10^{4}\,$Pa.nm (gel phase).
Inset: time-dependent diffusion coefficient $D(t)$ obtained from the corresponding master curves by Eq.~\ref{timedependentDt}.
}
\label{MSD}
\end{figure*}

	Figure~\ref{psi_beta} illustrates the behavior of the functions $\psi(t)$ and $\beta(t)$ that were obtained from the experimental data presented in Fig.~\ref{MSD}.
	Interestingly, both $\psi(t)$ and $\beta(t)$ display positive values and present the same general behavior for both sol and gel phases, even though the sol phase required negative values for $\alpha$ and $\kappa$.
	At short times, $t \ll \tau$, the logarithm in Eq.~\ref{psit} can be expanded around 1, so that $\psi(t) \approx (\alpha/2\kappa) (t/\tau)^{n}$ and $\beta(t) \approx (\alpha n/2\kappa \tau) (t/\tau)^{n-1}$.
	On the other hand, at later times, $t \gg \tau$, one have that $(t/\tau)^n + 1 \approx (t/\tau)^n$ so that Eqs.~\ref{psit} and~\ref{betat}  yield  $\psi(t) \approx (\alpha n/2\kappa) \ln (t/\tau)$ and  $\beta(t) \approx (\alpha n/2\kappa \tau) (t/\tau)^{-1}$, respectively.
	Contrary to what is observed for the corresponding memory function $\beta(t)$ evaluated from Eq.~\ref{MSDkrallweitz}, the limit of Eq.~\ref{betat} at long times, {\it i.e.},~$\beta(t) \propto t^{-1}$, is consistent with fluctuating hydrodynamics theory~\cite{adelman1976jcp}, and, accordingly, it does not depend either on $n$ or $\alpha$ (see Fig.~\ref{psi_beta}).

%%%%%%%%%%
%% FIG. 1: psi(t) and memory function \beta(t)
%%%%%%%%%%

%%%%%%%%%%
%% FIG. 2: MSD and shear modulus
%%%%%%%%%%

\subsection{Probe particle dynamics}
\label{probeparticledyn}

	Now, by considering the response function given by Eq.~\ref{generalized_chit}, one can obtain the expression for the MSD through Eq.~\ref{MSDpsit}, which gives,
\begin{equation}
\langle \Delta r^2 (t) \rangle 
= \frac{d k_BT}{\kappa}
\left\{1-
\left[
  \left(  \frac{t}{ \tau} \right)^{n} + 1 
\right]^{-\alpha}
\right\}
 ~~.
\label{rizzi_solution}
\end{equation}
	Importantly, such expression can be thought as obtained by averaging the trajectories over many mesoscopic regions of the sample just like in the experiments that use multi-particle techniques~\cite{rizzi2018eac}.
	Also, it can be considered an alternative to Eq.~\ref{MSDkrallweitz} and, as explored in Ref.~\cite{rizzi2017jcp}, it can be used to describe the dynamics of probe particles immersed in viscoelastic materials with a semisolid response like gels~\cite{raobook}.
	In fact, as I demonstrate below, Eq.~\ref{rizzi_solution} is a general expression that can be used to describe the MSD of probe particles immersed in a complex fluid, {\it i.e.},~in the sol phase, as well.

	In addition, one can also evaluate the time-dependent diffusion coefficient $D(t)$ from Eqs.~\ref{generalized_chit} and~\ref{betat} through Eq.~\ref{Dtmut}, which yields
\begin{equation}
D(t) = k_B T \frac{\alpha n}{2 \kappa \tau} \left( \frac{t}{\tau} \right)^{n-1}
\bigg/ ~
 \left[ \left( \frac{t}{\tau} \right)^{n} + 1 \right]^{1+\alpha} ~~.
\label{timedependentDt}
\end{equation}
	It is worth noting that, just as discussed earlier for the functions $\psi(t)$ (Eq.~\ref{psit}) and $\beta(t)$ (Eq.~\ref{betat}), $\alpha$ should adopt the same sign of $\kappa$ in order to have $D(t)$ as a positive-valued function.~Also, 
	as expected from Sec.~\ref{fokkerplanckEQ}, the long time behavior of the memory function, {\it i.e.},~$\beta(t) \approx (\alpha n/2 \kappa) t^{-1}$, is clearly different from that of Eq.~\ref{timedependentDt} when $t \gg \tau$, that is
\begin{equation}
D(t) \approx k_B T \frac{\alpha n}{2 \kappa \tau}  \left( \frac{t}{\tau} \right)^{-(1 + \alpha n)} ~~.
\label{Dt_long_times}
\end{equation}
	Interestingly, if one defines $\delta(t) = (2d)^{-1} \langle \Delta r^2 (t) \rangle / t$ with $\langle \Delta r^2 (t) \rangle$ given by Eq.~\ref{rizzi_solution} then, at later times ($t \gg \tau$), $\delta(t) \approx (k_BT/2\kappa) t^{-1}$ if $\kappa$ is positive.~If
	$\kappa$ is negative and $\alpha \approx - 1/n$, then one obtain basically
the same result, {\it i.e.},~$\delta(t) \approx - (k_BT/2\kappa) t^{\alpha n}$, with $\kappa<0$ and $\alpha n \approx -1$.

	In order to validate my approach, I include in Fig.~\ref{MSD} results that were obtained from the experimental data taken from Ref.~\cite{furst2008prl}, where the authors studied the sol-gel transition of chemically cross-linked polyacrylamide gels.~The
	microrheology experiments were done with the particle tracking videomicroscopy technique ($d=2$) using fluorescent polystyrene microspheres with radius $a=525\,$nm, and I presume that the experiments were done at room temperature, {\it i.e.},~$T=298\,$K (or $k_BT = 4.114\,$pN.nm).

	The master curves for the MSD were generated from several experiments at different concentrations ($c$ in \%wt) of bis-acrylamide cross-linker, and one can recover the actual data by using the shift factors $\text{a}_c$ and $\text{b}_c$ that are available in Ref.~\cite{furst2008prl}.
	Figure~\ref{MSD}(a) displays the typical behavior observed for probe particles immersed in the sol phase, with the long time diffusive behavior being characteristic of a fluid-like solution with $\langle \Delta r^{\,2}(t) \rangle \propto t$ and $D(t)$ 
 independent of time.~Figure~\ref{MSD}(b)
	 corresponds to a restricted motion of the probe particles, which is tipically observed in the gel phase, with $\langle \Delta r^{\,2}(t) \rangle$ going to a constant value for long times.
	Indeed, when $\alpha > 0$ (gel phase), Eq.~\ref{rizzi_solution} leads to
$\langle \Delta r^2 (t) \rangle \approx d k_BT/\kappa$ and $D(t)$ decays with $t^{-(1+\alpha n)}$ as in Eq.~\ref{Dt_long_times} for $t \gg \tau$.
	When $\alpha\approx -1/n$ (sol phase), Eq.~\ref{Dt_long_times} yields a diffusion coefficient
which is constant, $D_0 = - k_B T / (2 \kappa \tau)$, 
and a MSD given by $\langle \Delta r^2 (t) \rangle \propto t$.
	By assuming that, at long times, the particle in the sol phase probes an effective viscosity given by $\eta_0$, one can assume that $D_0= k_BT/6 \pi a \eta_0$, which is equivalent to
$\kappa = - 3 \pi a \, \eta_0 \, \tau^{-1}$,
where the negative value of $\kappa$ indicates that the effective elastic force is repulsive, with the corresponding potential being a barrier instead of a well~\cite{satija2019jphyschemA}.
	Importantly, this last result reveals that, at least for the sol phase, there is a direct link between the effective elastic constant $\kappa$ defined by in the Langevin equation, Eq.~\ref{GLE}, and a rheological property of the solution, {\it i.e.},~its viscosity $\eta_0$.

	As shown in Fig.~\ref{MSD}, the dynamics of probe particles both on sol and gel phases share the same short time behavior, with the MSD and the time-dependent diffusion coefficient given, respectively, by $\langle \Delta r^2 (t) \rangle \propto t^{n}$ and $D(t) \propto t^{n-1}$,
with the exponent $n=0.55$.
	Figure~\ref{psi_beta} shows that a similar short-time behavior is shared by the functions $\psi(t) $ and $\beta({t})$, respectively.
	The analogy with the response of polymer solutions presented in Sec.~\ref{polymer_solutions} indicates that, since $n$ is slightly above $1/2$, the clusters in the polyacrylamide solution might be not made exclusively of long flexible structures, 
but it might include semiflexible and short structures as well~\cite{leonam2020jnonnew}.

	Interestingly, by comparing the limiting case of the time-dependent diffusion coefficient given by Eq.~\ref{Dt_long_times} with $D(t) \propto t^{-d_s/2}$, one finds that 
$d_s = 2 (1 + \alpha n)$, where $d_s$ is an estimate for the spectral dimension of the network in the gel phase~\cite{teixeira2007jphyschemB}.
	For $n=0.55$ and $\alpha=0.71$ one finds $d_s = 2.78$, 
which is consistent with the values reported in Ref.~\cite{teixeira2007jphyschemB} for other kinds of gels.

\subsection{Viscoelastic response functions}

\begin{figure}[!t]
\centering
\includegraphics[width=0.43\textwidth]{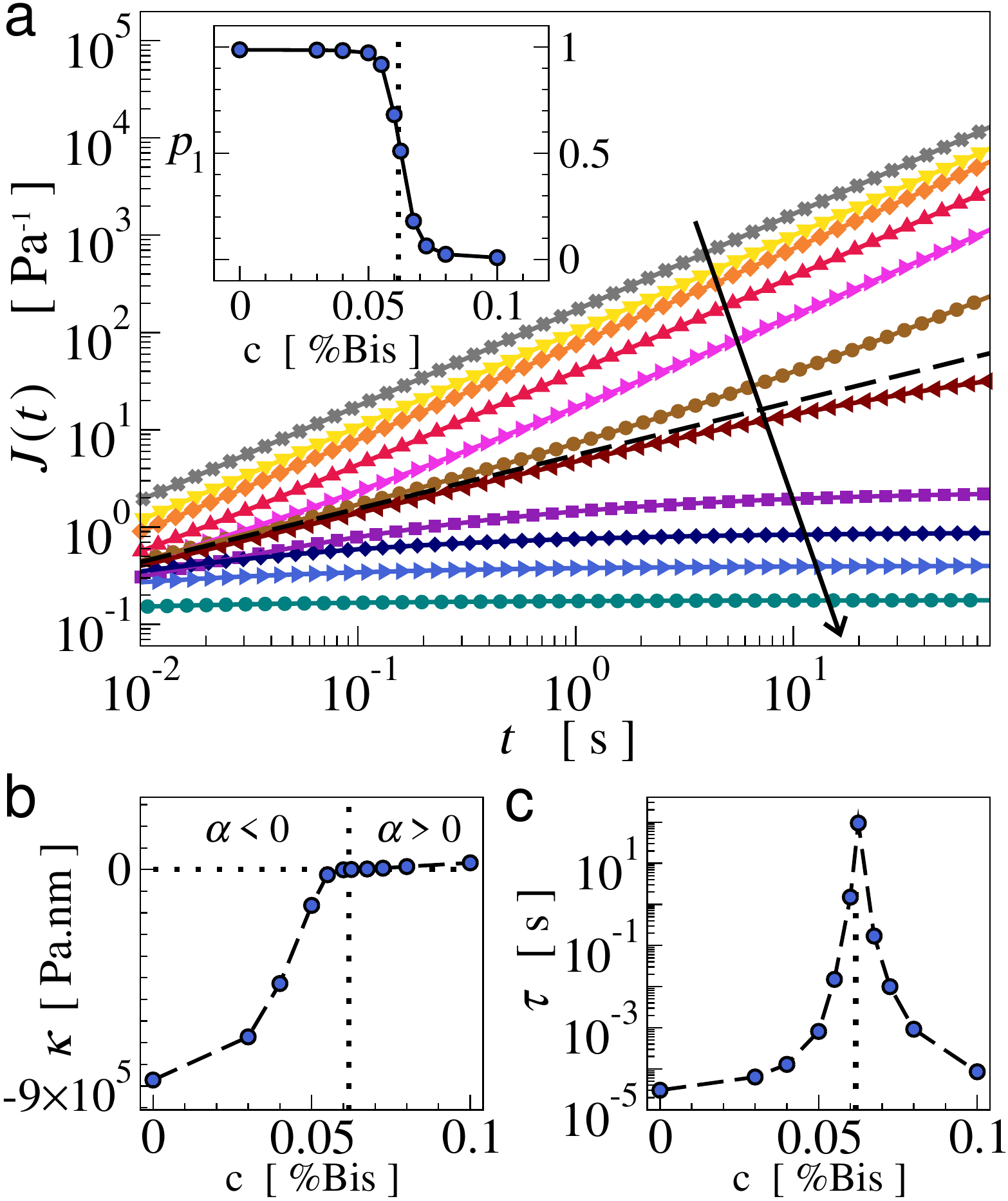}
\caption{(a) Compliance $J(t)$ for different concentrations $c$ of bis-acrylamide cross-linker.
	Symbols denote the data obtained from the master curves (see Fig.~\ref{MSD}) by using the experimentally obtained shift factors $\text{a}_c$ and $\text{b}_c$ of Ref.~\cite{furst2008prl}, while continuous lines correspond to the fit of the data points to Eq.~\ref{compliance}.
	The dashed line denote a curve proportional to $t^n$ with $n=0.55$, which lies between the sol and gel phases, 
and the arrow indicate the increasing concentration of cross-linker $c$.
	Inset shows the effective exponent of $J(t)$, Eq.~\ref{puexponent}, for 
different concentrations at $t=1\,$s.
(b) and (c) are, respectively, the elastic constant $\kappa$ and the relaxation time $\tau$ obtained from the fit of the
data displayed in (a) to Eq.~\ref{compliance} with the same exponents $\alpha$ and $n$ used in Fig.~\ref{MSD}.
	The vertical dotted lines indicate the critical concentration of cross-linkers $c^{*}$.
}
\label{complianceJt}
\end{figure}

	As indicated earlier, the complex shear modulus $G^{*}(\omega)$ of a viscoelastic material can be evaluated from the compliance $J(t)$ through Eq.~\ref{complex_modulus_compliance}.
	Also, by considering the MSD of probe particles defined by Eq.~\ref{rizzi_solution}, one can obtain the compliance from Eq.~\ref{compliance_MSD}, which yields
\begin{equation}
J(t) = \frac{3 \pi a}{\kappa} 
\left\{1-
\left[
  \left(  \frac{t}{ \tau} \right)^{n} + 1 
\right]^{-\alpha}
\right\}~~~.
\label{compliance}
\end{equation}
	From the master curves of the MSD presented in Figs.~\ref{MSD}(a) and (b), 
one can use the shift factors $\text{a}_c$ and $\text{b}_c$ in Ref.~\cite{furst2008prl} to obtain the estimates of the compliance $J(t)$ at different concentrations $c$ of bis-acrylamide, as shown in
Fig.~\ref{complianceJt}(a).
	Figures~\ref{complianceJt}(b) and~\ref{complianceJt}(c) show
the values of $\kappa$ and $\tau$ as a function of $c$, which were obtained
from a fit of the data presented in Fig.~\ref{complianceJt}(a) to Eq.~\ref{compliance}.
	Although both parameters display a non-linear behavior as 
a function of the cross-linker concentration, the compliance curves $J(t)$ goes down
as $c$ increases, indicating that the formation of structures is impinging a more restricted movement to the probe particles.
	Interestingly, the relaxation time $\tau$ shows a divergent-like behavior which is typical of the sol-gel transition, while $\kappa$ displays a non-linear behavior with negative values for $c<c^*$ when $\alpha<0$, and  positive values for $c>c^*$, with~\cite{furst2008prl} $c^{*}\approx0.0617\%$.

%%%%%%%%%%
%% FIG. 3: Compliance J(t)
%%%%%%%%%%

%%%%%%%%%%
%% FIG. 4: Shear modulus, Viscosity
%%%%%%%%%%
\begin{figure*}[!t]
\centering
\includegraphics[width=0.90\textwidth]{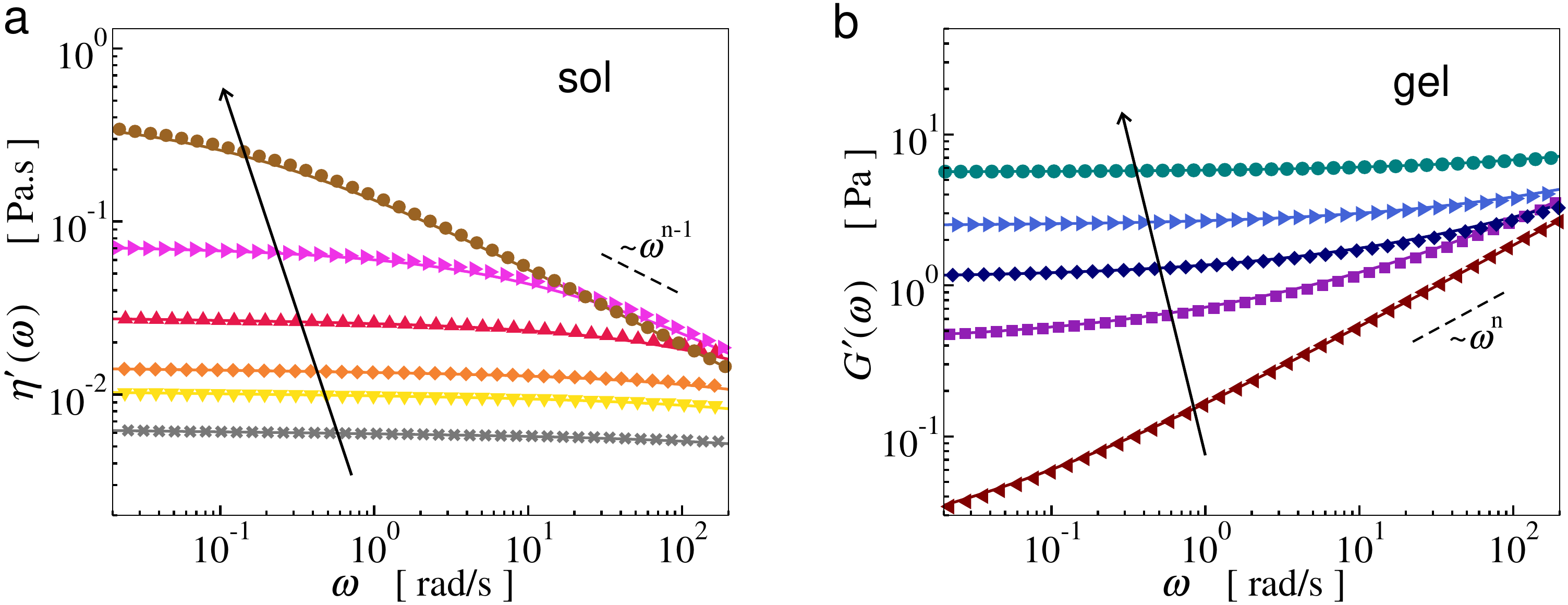}
\caption{(a) and (b) show the viscosity $\eta^{\prime}(\omega)=G^{\prime \prime}(\omega)/\omega$ 
and the storage modulus $G^{\prime}(\omega)$ at different concentrations $c$ of cross-linkers, respectively.
	Arrows indicate the increasing concentration $c$ of cross-linkers.
	Symbols correspond to data obtained numerically 
from the curves $J(t)$ presented in Fig.~\ref{complianceJt}
by using the method proposed in Ref.~\cite{manlio2009pre}, 
while continuous lines corresponds to Eqs.~\ref{gprimep}, $G^{\prime}(\omega)$, and~\ref{viscosityprimep}, $\eta^{\prime}(\omega)$, 
with the exponents $\alpha$ and $n$ obtained from Fig.~\ref{MSD}, and the values of $\kappa$ and $\tau$ extracted from the data presented in Figs.~\ref{complianceJt}(b) and~\ref{complianceJt}(c), respectively.
	For both $\eta^{\prime}(\omega)$ and $G^{\prime}(\omega)$ the relative difference between the estimates obtained by the two approaches is less than 0.1.
	Dashed lines indicate the behaviors at high frequencies: 
$\eta^{\prime}(\omega) \propto \omega^{n-1}$ and 
$G^{\prime}(\omega) \propto \omega^{n}$.
}
\label{shearmodulus}
\end{figure*}

	Unfortunately, it is difficult to evaluate an exact and general expression ({\it i.e.},~for any values of $n$, $\alpha$, $\tau$ and $\kappa$) for the Fourier-Laplace transform of the Eq.~\ref{compliance}.
	However, one can obtain approximated results by assuming that
the compliance can be expanded around a time $t=1/\omega$ as~\cite{mason2000rheolacta}
\begin{equation}
J(t) \approx
J( 1/\omega ) (\omega t)^{p(\omega)}
\label{compliance_approx}
\end{equation}
where $p(\omega) \equiv p(t)|_{t=1/\omega}$ is an effective exponent of the compliance (or, equivalently, of the MSD) around a time $t=1/\omega$, with $p(t)$ given by
\begin{equation}
p(t) = \frac{d \ln \langle \Delta r^2(t) \rangle }{d \ln t}
= 2d \frac{D(t)}{\langle \Delta r^2(t) \rangle }  t ~~.
\label{exponent_pt}
\end{equation}
	By considering $\langle \Delta r^2(t) \rangle$ and $D(t)$ given by Eqs.~\ref{rizzi_solution} and~\ref{timedependentDt}, respectively,~Eq.~\ref{exponent_pt} yields
\begin{equation}
p(\upsilon) = \alpha n  \frac{(\upsilon^{-1} -1 )}{(1 - \upsilon^{\alpha})}  ~~,
\label{puexponent}
\end{equation}
with $\upsilon = \upsilon(t) = 1 + (t/\tau)^n$.
	One can check that, at short times ($t \ll \tau$), $\upsilon \approx 1$, thus
$p = n$ for both $\alpha >0$ and $\alpha < 0$, % via L'Hopital!!!!
which is consistent to what is observed for the MSD in Figs.~\ref{MSD}(a) and ~\ref{MSD}(b), and for $J(t)$ in Fig.~\ref{shearmodulus}(a).
	For long times ($t \gg \tau$), $\upsilon \approx (t/\tau)^n$ so that
$p = -\alpha n \approx 1$ for $\alpha < 0$ (sol phase), and $p(t) = \alpha n (t/\tau)^{-\alpha n} \approx 0$ for $\alpha > 0$ (gel phase), in agreement to Figs.~\ref{MSD}(a) and~\ref{MSD}(b), and Fig.~\ref{shearmodulus}(a) as well.
	As illustrated in the inset of Fig.~\ref{complianceJt}(a), one can use Eq.~\ref{puexponent} to obtain the exponent $p_1$ at given time, {\it e.g.},~$t=1\,$s, for the different values of $c$ in order to estimate the gelation concentration $c^{*}$ (see, {\it e.g.},~Ref.~\cite{larsen2009macromol}).

	Hence, one can consider Eq.~\ref{compliance_approx} to obtain
an estimate of $\hat{J}(\omega)$ through a Fourier-Laplace integral,
{\it i.e.},~$\hat{J}(\omega) \approx J(1/\omega) \mathcal{L}[(\omega t)^{p(\omega)};s]_{s=i\omega}$, which leads to
\begin{equation}
i \omega \hat{J}(\omega) \approx
J(1/\omega) \, \Gamma(1 + p(\omega)) \, e^{- i \pi p(\omega) /2}~~.
\label{Fucompliance}
\end{equation}
	As previously mentioned,~$\Gamma(1 + p(\omega))$ denotes the usual gamma function~\cite{gradshteynbook}.
	The above expression is valid for $p(\omega)>-1$, so it can be used to describe all types of diffusive regimes, including the subdiffusion observed in restricted random walks where $p \approx 0$.
	Finally, by considering Eq.~\ref{Fucompliance}, one can use Eq.~\ref{complex_modulus_compliance} to evaluate the storage modulus $G^{\prime}(\omega)$ and the viscosity 
$\eta^{\prime}(\omega) = G^{\prime \prime}(\omega) / \omega$, which are given, respectively, as
\begin{equation}
G^{\prime}(\omega) \approx g(\omega) \left( \frac{\kappa}{3 \pi a} \right) \cos \left[ \frac{\pi p(\omega)}{2} \right] ~~,
\label{gprimep}
\end{equation}
and
\begin{equation}
\eta^{\prime}(\omega)  \approx \frac{g(\omega)}{\omega} \left( \frac{\kappa}{3 \pi a} \right) \sin \left[ \frac{\pi p(\omega)}{2} \right]  ~~,
\label{viscosityprimep}
\end{equation}
where
\begin{equation}
g(\omega) = \frac{1 }{ \{ 1 - [ (\omega \tau)^{-n} + 1]^{-\alpha} \} \Gamma(1 + p(\omega))}
\label{gomega}
\end{equation}
with $p(\omega)$ obtained from Eq.~\ref{puexponent} assuming that $\upsilon(\omega)=\upsilon(t)|_{t=1/\omega}=1 + (\omega \tau)^{-n}$.

	Figures~\ref{shearmodulus}(a) and~\ref{shearmodulus}(b) include a comparison between 
the viscosity $\eta^{\prime}(\omega)$ (at the sol phase) and the storage modulus $G^{\prime}(\omega)$ (at the gel phase) evaluated from Eqs.~\ref{gprimep} and~\ref{viscosityprimep}, and computed directly from the compliance $J(t)$ presented in Fig.~\ref{complianceJt}(a) via the numerical method proposed in Ref.~\cite{manlio2009pre}.
	At high frequencies, $\omega \gg \tau^{-1}$, one have that $p \approx n$ and $g(\omega) \approx \alpha^{-1} (\omega \tau)^n / \Gamma(1+n)$, so that, just as discussed in Sec.~\ref{polymer_solutions}, one have that $G^{\prime}(\omega) \propto \omega^{n}$ and $G^{\prime \prime}(\omega) \propto \omega^{n}$, which leads to $\eta^{\prime}(\omega) \propto \omega^{n-1}$, as shown in Fig.~\ref{shearmodulus}.

	For the sol phase ($\alpha<0$) at low frequencies, $\omega_0 \ll \tau^{-1}$, Eqs.~\ref{puexponent} and~\ref{gomega} lead to $p(\omega_0) = - \alpha n$ and $g(\omega_0) \approx - (\omega_0 \tau)^{-\alpha n}$, respectively.
	Thus, one can consider Eq.~\ref{viscosityprimep} to obtain
a limit for the viscosity at low frequencies, that is
\begin{equation}
\eta^{\prime}(\omega_0) = - \frac{\kappa}{3 \pi a} \, \omega_0^{-(1+\alpha n)} \, \tau^{-\alpha n} ~~.
\label{lowfreqetaprime0sol}
\end{equation}
	By taking $\alpha n \approx -1$, the above equation yields a value $\eta_0=\eta^{\prime}(\omega_0)$ which is practically independent of the frequency $\omega_0$, as shown in Fig.~\ref{shearmodulus}(a).
	Also, it yields
\begin{equation}
\kappa \approx - 3 \pi a \frac{\eta_0}{ \tau}~~,
\label{kappaeta0tau}
\end{equation}
which corresponds to the same relation obtained from the limit for long times of the time-dependent diffusion coefficient $D(t)$ in Sec.~\ref{probeparticledyn}.
	In addition, at the sol phase, the low frequency value of the storage modulus given by Eq.~\ref{gprimep} can be estimated as
\begin{equation}
G^{\prime}(\omega_0) 
\approx \, C_{s} \, \eta_0 \,  \omega_0 ~~,
\label{lowfreqGprime0sol}
\end{equation}
where $\eta_0$ is defined as in Eq.~\ref{kappaeta0tau}, and
 $C_{s} = \pi(1+\alpha n) /2$ with $\alpha n \approx -1$.
	As expected, for $\alpha<0$ (sol phase), $G^{\prime}(\omega_0)$ is proportional to $\omega_0$, which is a characteristic behavior of complex solutions at low frequencies.

%%%%%%%%%%
%% FIG. 5: 
%%%%%%%%%%
\begin{figure}[!t]
\centering
\includegraphics[width=0.46\textwidth]{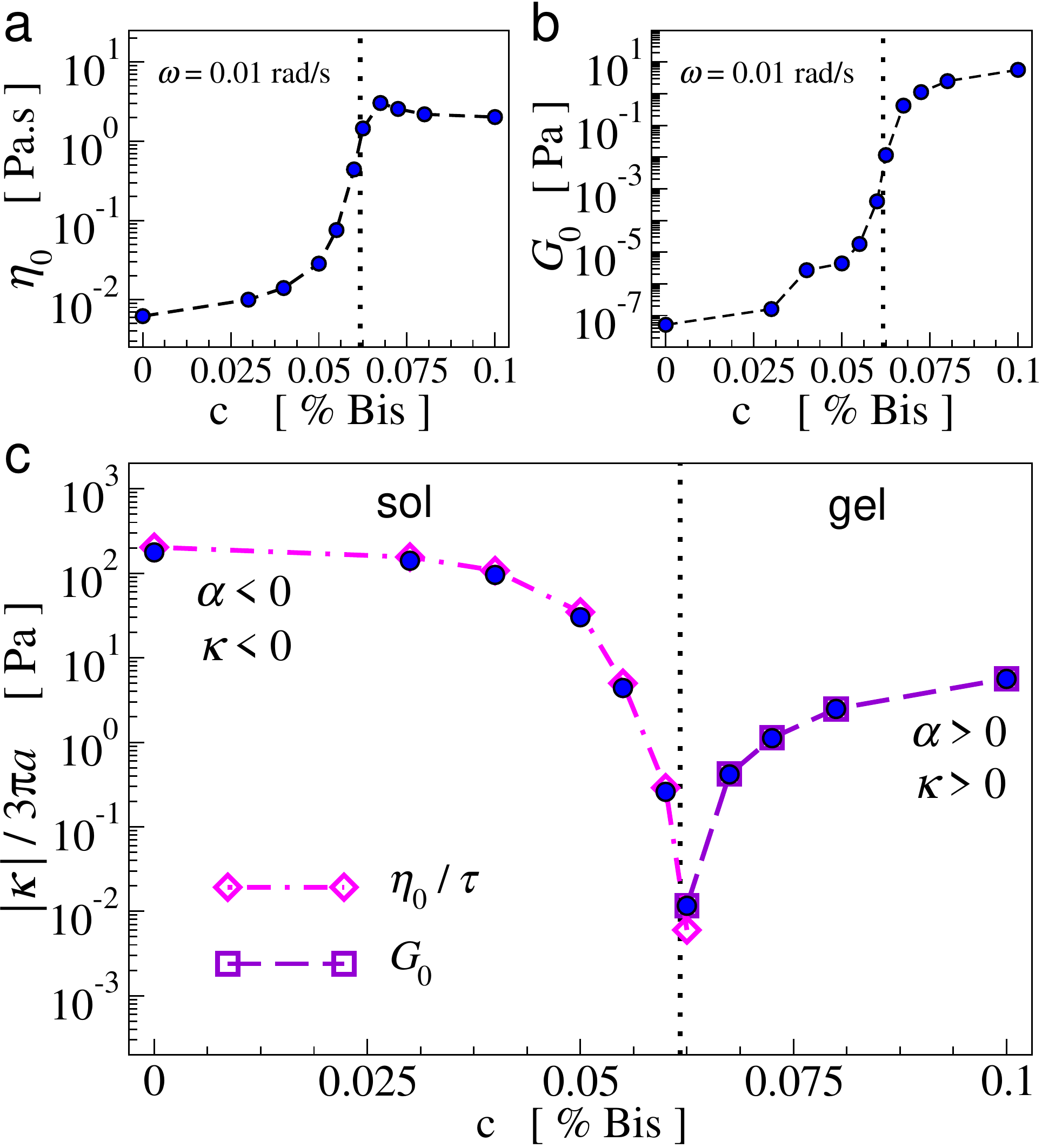}
\caption{
(a) and (b) show,
respectively, the viscosity $\eta^{\prime}(\omega_0)$ and the storage modulus $G^{\prime}(\omega_0)$ obtained from the data presented in Figs.~\ref{shearmodulus}(a) and~\ref{shearmodulus}(b) at a low frequency, $\omega_0=0.01\,$rad/s, for different concentrations of cross-linkers $c$.
(c) Validation of Eqs.~\ref{kappaeta0tau} and~\ref{kappaG0}: filled circles correspond to the modulus of the effective elastic constant, {\it i.e.}~$|\kappa|/3 \pi a$; the values of $\eta_0$ and $G_0$ are those displayed in (a) and (b), respectively, while the value of $\tau$ is obtained from Fig.~\ref{complianceJt}(c).
	The vertical dotted lines indicate the critical concentration of cross-linkers $c^{*}$.
}
\label{kappaeta0G0relationship}
\end{figure}

	At the gel phase, one have that $\alpha>0$, then, at low frequencies ($\omega_0 \ll \tau^{-1}$),
Eqs.~\ref{puexponent} and~\ref{gomega} lead to $p(\omega_0) = \alpha n (\omega_0 \tau)^{\alpha n}$ and $g(\omega_0) \approx 1$, respectively.
	Hence, one can use Eq.~\ref{gprimep} to obtain the limit of the storage modulus at low frequencies, that is
\begin{equation}
G^{\prime}(\omega_0) \approx \frac{ \kappa }{ 3\pi a}  ~~,
\label{lowfreqGprime0gel}
\end{equation}
which corresponds to the plateau value showed in Fig.~\ref{shearmodulus}(b).
	And, by assuming that $G_0 \equiv G^{\prime}(\omega_0)$ at the gel phase, one have that
\begin{equation}
\kappa \approx 3 \pi a \, G_0  ~~.
\label{kappaG0}
\end{equation}
	Also, at low frequencies and $\alpha>0$, Eq.~\ref{viscosityprimep} yields
\begin{equation}
\eta^{\prime}(\omega_0) = C_g \, G_0  \, \tau^{\alpha n}  \, \omega_0^{-(1-\alpha n)}~~,
\label{lowfreqetaprime0gel}
\end{equation}
where $G_0$ is defined by Eq.~\ref{kappaG0} and $C_{g}= \alpha n (\pi / 2)$.

	Figures~\ref{kappaeta0G0relationship}(a) and~\ref{kappaeta0G0relationship}(b) show, respectively, $\eta^{\prime}(\omega_0)$ and $G^{\prime}(\omega_0)$ obtained from the data 
presented in Figs.~\ref{shearmodulus}(a) and~\ref{shearmodulus}(b) at a constant low frequency, $\omega_0=0.01\,$rad/s, as a function of the concentration of cross-linkers $c$ through the sol-gel transition for the polyacrylamide.
	At the sol phase, the low frequency values of the viscosity and the storage modulus are given by Eqs.~\ref{lowfreqetaprime0sol} and \ref{lowfreqGprime0sol}, respectively.
	At the gel phase, the storage modulus can be described by Eq.~\ref{lowfreqGprime0gel}, while
the viscosity is given by Eq.~\ref{lowfreqetaprime0gel}.

	Finally, I emphasize that Eqs.~\ref{kappaeta0tau} and~\ref{kappaG0} establish relationships between the effective elastic constant $\kappa$ defined in the Langevin equation, Eq.~\ref{GLE}, and the rheological properties of the viscoelastic material, {\it i.e.}~$\eta_0/\tau$ at the sol phase, and $G_0$ at the gel phase.
	Figure~\ref{kappaeta0G0relationship}(c) illustrates this correspondence through the sol-gel transition, by presenting the values of the modulus of the effective elastic constant, $|\kappa|/3 \pi a$, together with the values of $\eta_0 / \tau$ and $G_0$ obtained from the data in Figs.~\ref{shearmodulus}(a) and~\ref{shearmodulus}(b).

\section{Concluding remarks}

	In this paper I consider an approach based on microrheology to provide a general expression for the mean-squared displacement of probe particles which can be used to obtain the full viscoelastic response of gels.
	The non-markovian Langevin approach used here allowed me to establish useful interrelations between the MSD of probe particles, Eq.~\ref{rizzi_solution}, and several functions, in particular, the response function $\chi(t)$, Eq.~\ref{generalized_chit}, the time-dependent diffusion coefficient $D(t)$, Eq.~\ref{timedependentDt},
the memory kernels $b(t)$ and $u(t)$ of the Langevin equations, Eqs.~\ref{GLE} and~\ref{GLE_alt}, and also to the memory functions $\psi(t)$,  Eq.~\ref{psit}, and $\beta(t)$,  Eq.~\ref{betat}, which are related to the Fokker-Plank equation, Eq.~\ref{fokker_planck_eq}.

	The expressions obtained for the viscoelastic response of the gel, {\it i.e.},~the storage modulus, Eq.~\ref{gprimep}, and the viscosity, Eq.~\ref{viscosityprimep}, should be of practical interest to both theoreticians and experimentalists.
	For instance, those equations might be used to describe rheological/microrheological data providing estimates for the exponents $n$ and $\alpha$ which characterize, respectively, the dynamics and the structures of the gel.
	In particular, Eq.~\ref{viscosityprimep} can be seen as an alternative model ({\it e.g.},~Cross, Carreau, Bird) to viscosity, specially for shear-thinning viscoelastic materials, which present a decreasing $\eta^{\prime}(\omega)$ similar as those displayed in Fig.~\ref{shearmodulus}(a).
	Also, the analytical expressions which link the effective elastic constant $\kappa$ introduced in the Langevin equations to the viscoelastic properties of the gel, {\it i.e.}~Eqs.~\ref{kappaeta0tau} and~\ref{lowfreqGprime0sol} (sol phase), and~Eqs.~\ref{kappaG0} and~\ref{lowfreqetaprime0gel} (gel phase), should provide a theoretical basis that goes beyond the analogy between gelation and percolation~\cite{martin1990macromol,martin1991review}, since those equations are expected to be valid not only at the sol-gel transition but in the coarsening regime as well~\cite{rizzi2017jcp,rizzi2016sm,rizzi2015prl}.

	It is worth mentioning that, due to experimental resolution of the available data, I consider that the exponent $\alpha$ changes abruptly from $-1/n$ to a finite positive value at the gelation point, 
even though the effective exponent $p_1$ changes continuosly (see the inset of Fig.~\ref{complianceJt}(a)).
	However, for time-cure experiments involving physical gels, it could be 
that the changes in $\alpha$ are more notiaceable, and one might observe what is happening with it at the gelation point. 
	In that case, just like in any microrheology experiment~\cite{rizzi2018eac}, ergodicity breaking should be avoided by considering that the observation times $t_{\text{e}}$, i.e., the longest times used to probe the MSD, are much shorter than the cure times $t_{\text{w}}$.
	Hence, in order to use the framework developed here, one must assume that the distribution $\rho_{\alpha}(\varepsilon)$ defined by Eq.~\ref{k_distribution} remains unaltered for times $t< t_{\text{e}} \ll t_{\text{w}}$.

	Finally, one should note that I derived the expression for the MSD from a distribution of elastic constants which might be related to the cluster sizes distribution in the sample~\cite{krall1997physA,dinsmore2006prl}.~Thus, 
	my results suggest that, besides of being interpreted as a generalization of the results obtained in Ref.~\cite{krall1997physA}, the expression obtained here for the MSD, Eq.~\ref{rizzi_solution}, should be related to approaches that compute the relaxation modulus $G(t)$ from the time relaxation distribution $H(\lambda)$ through Eq.~\ref{relaxation_modulus} as in, for instance, Ref.~\cite{winter1986jrheol}.
	In fact, by assuming that $\xi = (\lambda/\tau)^{\gamma}$ and recalling that $\xi=\alpha \varepsilon/2 \kappa$, one can consider the distribution of local elastic constants $\varepsilon$ defined by Eq.~\ref{k_distribution} to obtain the distribution of relaxation times $\lambda$ as
\begin{equation}
H(\lambda) = 
\frac{1}{\, \Gamma(\alpha/2)}
\left|\frac{\gamma}{\tau} \right|
\left( \frac{\lambda}{\tau} \right)^{\alpha\gamma/2 - 1}
\exp \left[ - \left(\frac{\lambda}{\tau}  \right)^{\gamma} \, \right]
~~,
\label{lambda_distribution}
\end{equation}
which is known as a generalized gamma distribution~\cite{crooksbook}.~Remarkably, 
	the above expression is similar to the semi-empirical distribution used in Ref.~\cite{zaccone2014jrheol} to obtain the complex modulus $G^{*}(\omega)$ of liquid-like solutions of colloidal particles from their self-assembly aggregation kinetics, and it is worth mentioning that one might explore it also to describe gel-like responses by considering negatives values for $\gamma$.

\section*{AcknowledgmentS}

	The author acknowledge useful discussions with Stefan Auer, David Head, Manlio Tassieri, and \'Alvaro Teixeira, and the financial support of the brazilian agency CNPq (Grants 
N\textsuperscript{o}  306302/2018-7 
% Bolsas de Produtividade em Pesquisa - PQ
and 
N\textsuperscript{o} 426570/2018-9). 
% Chamada MCTIC/CNPq N 28/2018 - Universal/Faixa A - Ate R$ 30.000,00
	FAPEMIG (Process APQ-02783-18) is also acknowledged, although no funding was released until the submission of the present work.

%\section*{Bibliography}

%\bibliographystyle{elsarticle-num}
%\bibliography{biblio}{}


\end{document}